\documentclass[conference]{IEEEtran}
\IEEEoverridecommandlockouts
\usepackage{cite}
\usepackage{amsmath,amssymb,amsfonts}
\usepackage{listings}
\usepackage{graphicx}
\usepackage{textcomp}
\usepackage{xcolor}
\usepackage{color}
\usepackage{hyperref}
\def\BibTeX{{\rm B\kern-.05em{\sc i\kern-.025em b}\kern-.08em
    T\kern-.1667em\lower.7ex\hbox{E}\kern-.125emX}}

\definecolor{mygreen}{rgb}{0,0.6,0}
\definecolor{mygray}{rgb}{0.5,0.5,0.5}
\definecolor{mymauve}{rgb}{0.58,0,0.82}
\begin{document}

\title{Elastic execution of checkpointed MPI applications}

\author{
\IEEEauthorblockN{Sumeet Gajjar}
\IEEEauthorblockA{\textit{Khoury College of Computer Sciences} \\
\textit{Northeastern University}\\
Boston, MA\\
gajjar.s@northeastern.edu}
\and
\IEEEauthorblockN{Saurabh Vaidya}
\IEEEauthorblockA{\textit{Khoury College of Computer Sciences} \\
\textit{Northeastern University}\\
Boston, MA\\
vaidya.saur@northeastern.edu}
}

\maketitle


\begin{abstract}
MPI applications begin with a fixed number of rank and, by default, the rank remains constant throughout the application's lifetime. The developer can choose to increase the rank by dynamically spawning MPI processes. However doing this manually adds complexity to the MPI application. Making the MPI applications malleable \cite{b20} would allow HPC applications to have the same elasticity as that of cloud applications. We propose multiple approaches to change the rank of an MPI program agnostic to the modification of the user code. We use checkpointing as a tool to achieve mutability of rank by halting the execution and resuming the MPI program with a new state. In this paper, we focus on the scenario of increasing the rank of an MPI program using ExaMPI as the implementation for MPI.

\end{abstract}


\section{Introduction}
Message Passing remains a dominant programming model for distributed memory systems. Message Passing Interface (MPI) \cite{b1} is the de facto method for developing applications in scientific computing. The applications rely on MPI to achieve horizontal scaling with multiple nodes in HPC clusters. This kind of scaling is necessary to support memory and performance requirements.

The MPI applications running on HPC clusters have mostly been designed to use a fixed number of resources. Thus, the application cannot exploit elasticity when additional resources are available. MPI versions 1.x did not have any support for changing the number of processes during the execution. MPI 2.0 introduced a feature to increase the number of processes running under MPI using MPI\_Comm\_spawn \cite{b2} and MPI\_Comm\_spawn\_multiple \cite{b3}. However, this feature is not supported by many of the available MPI implementations. The developer must put in extra effort to manage the newly spawned processes and redistribute the data efficiently amongst those new processes. 

 
Elastic execution of MPI applications is the process of modifying the allocated resources to the MPI application thereby increasing or decreasing the number of MPI processes running under the hood. There are a couple of cases that the elastic execution might help. One, for a long-running application, a user may want to increase the allocated resources to try and reduce the completion time of the application. Secondly, when an application is not scaling linearly, the number of instances can be reduced, resulting in lower nodes×hours, and thus a lower cost. 

In this paper, we propose three approaches to allow elastic execution of MPI applications. All of these approaches are implemented in userspace. Approaches A and C require modification to the underlying MPI library, while Approach B leverages the existing MPI APIs to achieve elasticity, with a caveat of modifying the MPI library to implement the corresponding APIs.

\section{Related work}
There are several research efforts related to elastic execution models. The literature for elastic execution models mostly relates to grid and cloud computing. The requirements for synchronization are low when it comes to grid and cloud computing as compared to scientific applications.

Raveendran et. al. \cite{b4} propose an approach to elastically execute existing MPI applications in the cloud. Nodes can be added and removed on the fly without hindering the current execution of the MPI application. To demonstrate this idea, the authors had to manually modify existing MPI applications to persist data at every iteration. However, this approach requires the MPI application to persist its state. The authors argue that this can be easily automated using a source-to-source translation tool.


Comprés et. al. \cite{b5} demonstrate the idea of elastic execution by proposing an MPI extension. The authors introduce four new MPI operations to achieve this. As proof of concept, the authors modified the MPI library --- MPICH \cite{b6} --- and the resource manager --- SLURM. \cite{b7}

\section{Background}
\subsection{DMTCP}

Modern supercomputers comprise thousands of nodes, and this number is constantly increasing. With the increasing number of hardware components in the supercomputers, a hardware fault is likely to occur sooner or later on these man-made marvels. The applications are however not developed to accommodate these hardware faults. Fault tolerance mechanisms can be added to the application; however, this is not the primary objective of the developer building a scientific application.

Checkpointing \cite{b8} is a technique that provides fault tolerance for computing systems. It consists of saving a snapshot of the application's state, so that the application can restart from that point in case of failure. This is particularly important for long-running applications that are executed in failure-prone computing systems. DMTCP --- Distributed Multi Threaded Checkpointing \cite{b9} --- is a user space library to transparently checkpoint applications. In this paper, we use DMTCP as our primary tool to checkpoint and restore an MPI process or the entire MPI application as a whole.

\subsection{DMTCP plugins}
DMTCP plugins \cite{b10} provide the flexibility to extend DMTCP generic library and write an add-on library. The user can use the DMTCP plugins to modify the state or any internal data structures of the program during, before or after checkpointing. In this paper we use the event hooks functionality of DMTCP plugins to capture various events generated by DMTCP and manipulate the data-structures to suit our needs.

\subsection{ExaMPI}
ExaMPI \cite{b11} is one of the implementations of the MPI standard developed with the goal of enabling researchers to
experiment rapidly and easily with its internals. It supports a modern MPI-3.x subset with a robust MPI-4.x road-map.

For the purpose of this paper, we need to understand the MPI runtime viz. mpiexec. mpiexec is a Python program which
takes the parameters to start the MPI application. mpiexec spawns multiple \emph{fault\_daemons}, one for each rank, with
fairly similar but different environment. \emph{fault\_daemons} are the wrappers which execute the actual MPI process.
Each \emph{fault\_daemon }with local-id 0 launches a \emph{head\_daemon}. Each node has one \emph{head\_daemon} which is responsible for
managing all the \emph{fault\_daemons} on that node. After launching the \emph{head\_daemon}, if the rank of the \emph{fault\_daemon }is 0
then a \emph{controller\_daemon} is launched. The \emph{controller\_daemon} is responsible for managing the entire MPI cluster.
Once the \emph{controller\_daemon} is launched, all the \emph{fault\_daemons} try to establish a communication link with the
respective \emph{head\_daemon}s. On success, the \emph{fault\_daemon }now commences the user defined MPI program and the MPI
cluster is initialized and ready for message passing.
\section{Proposed Solution}
\subsection{Approach A}
In the current design of ExaMPI, the only way to spawn a \emph{fault\_daemon }is using mpiexec. Since mpiexec exits after spawning the specified number of \emph{fault\_daemons}, there is no clean way to spawn additional \emph{fault\_daemons} for the same MPI application further in time.

We propose to change the design of the mpiexec and modify the order in which various daemons are launched. According to the proposed design, mpiexec launches the \emph{head\_daemons} first. The \emph{head\_daemon }spawns the \emph{fault\_daemons} which in turn starts MPI processes. The \emph{head\_daemon }with rank 0 spawns the \emph{controller\_daemon}. Because of this change in the ordering, we now have the ability to spawn \emph{fault\_daemons} without using mpiexec. We can signal \emph{controller\_daemon} to spawn new MPI processes. The \emph{controller\_daemon} based on the load balancing scheme will signal the corresponding \emph{head\_daemons}. The \emph{head\_daemons} spawns the respective number of \emph{fault\_daemons} which finally spawn the MPI processes.


The modified design of ExaMPI is needed to adapt to launching under DMTCP.
We launch the MPI application under DMTCP with --ckpt-open-files --allow-file-overwrite parameters --with-plugin exampi\_dmtcp\_plugin.so.
The --ckpt-open-files flag allows user to checkpoint the open files. The ExaMPI library has config files for each MPI process. The config file contains the information to establish TCP/UDP connection with any of the MPI ranks. These config files are created during the initialization of \emph{fault\_daemons}. As per the current implementation these config files are deleted once the application exits gracefully or if the application is interrupted in any manner for e.g. CTRL+C. The --ckpt-open-files flag also enables the restore of these deleted temporary files during the restart phase.


A further problem in the integration of ExaMPI with DMTCP is that
ExaMPI also creates files for logging purposes. DMTCP must checkpoint these files and try to restore them during the restart phase. But unlike config files the log files are available during the restart phase and when DMTCP tries restoring them, it encounters an error as the files already exists and hence cannot be overwritten. Specifying --allow-file-overwrite parameters flag resolves the issue and allows successful restart of the checkpointed MPI application.


During the init phase of DMTCP, the control is transferred momentarily to the exampi\_dmtcp\_plugin and a thread is created which waits on a conditional\_variable. After the thread is created, the control is returned back to DMTCP, and it continues the execution of the MPI application. During the restart phase, DMTCP restores the MPI application in memory and gives the control to the exampi\_dmtcp\_plugin. The conditional\_variable is signaled by the plugin and the control is returned back to DMTCP which resumes the execution of the MPI application.

The waiting thread now tries to establish a TCP connection with the \emph{controller\_daemon} in the ExaMPI runtime. Once the connection is established, a message is sent to the \emph{controller\_daemon} to increase the rank of MPI application. The \emph{controller\_daemon} now determines how many more MPI processes are required and the corresponding nodes to spawn the MPI processes. The controller now sends a message to the respective \emph{head\_daemons} to spawn \emph{fault\_daemons}. The \emph{head\_daemons} now spawns the \emph{fault\_daemons} which in turn spawn the MPI processes. The newly spawned processes are now blocked on the barrier \cite{b12} in MPI\_Init \cite{b13}.

During the restart phase, connection information of the newly spawned \emph{fault\_daemons} needs to be provided to the old MPI processes. This connection information and metadata in stored in config files and environment variables. The metadata information stored in the environment variables e.g. "EXAMPI\_WORLD\_SIZE" also needs to be updated. ExaMPI also uses "epoch" files to facilitate checkpoint and rollback to a previous state. However this feature is still in development stage and not included in the production release. Hence we are ignoring it in our design.

The old MPI processes now receive a error code MPI\_WORLD\_RESIZED on executing any of the MPI functions. The error code is a way to notify the already running MPI processes that the size of the world has changed. The processes can decide to re-partition data and distribute the work amongst the old as well as the newly spawned processes and make a call to enter the barrier. Thus, our new cluster will now have an increased rank.

Once the cluster upgrade is complete the new MPI processes are added to the MPI\_COMM\_WORLD communicator. No new communicator is created, the MPI\_COMM\_WORLD acts a unified communicator for old and new MPI processes. 


In addition to the approach described previously, there can be another alternative.  The alternative approach is to convert the MPI\_COMM\_WORLD into a versioned communicator. Initially the MPI\_COMM\_WORLD has the tag $v0$ And with each cluster upgrade the communicator expands but has a version tag associated with it. Let's consider a scenario where the MPI cluster was upgraded twice. For the upgrade from $x \xrightarrow{} y$ where $(y > x)$, the communicator gets the tag $v1$ and for the upgrade from $y \xrightarrow{} z$ where $(z > y)$, the communicator gets the tag $v2$. The communicator tag can be used to identify the membership of a given MPI process in a communicator at any given point in time. The MPI process can choose to use the tag along with the communicator to communicate with another MPI process. If the tag is not specified then the tag is defaulted to the latest one.

\begin{lstlisting}[caption=Code Snippet for Approach A, float]
                   
void repartition_data(int status_code){

	if (status_code == MPI_WORLD_RESIZED){
		/* code to repartition 
		 * or redistribute the data
		 */
		 
		MPI_barrier();
	}
}

int main(){

	//...
	int i, status_code;
	for (i = 0; i < TOTAL_ITERATIONS; i++)
	{
		status_code = MPI_Send();
		repartition_data(status_code);

		status_code = MPI_Recv();
		repartition_data(status_code);

		/* after any MPI operation, 
		 * repartition_data should be 
		 * called with the status code 
		 * of the operation to check 
		 * if data repartition is required
		 */
	}
	//...
}
\end{lstlisting}

\subsection{Approach B}

In this approach we use MPI\_Comm\_spawn and MPI\_Intercomm\_merge to launch new MPI processes and create a new communicator MPI\_COMM\_RESIZED\_WORLD which represents the resized world after restarting the checkpointed MPI application. 

We launch the MPI application under DMTCP in a similar manner as described in Approach A. The plugin however functions differently. During the plugin init phase we initialize the global variable MPI\_COMM\_RESIZED\_WORLD to MPI\_COMM\_NULL. MPI\_COMM\_NULL indicates the communicator is invalid and it is semantically correct since there is no resized world when the MPI application starts. A thread is also created which waits on a conditional\_variable. After the thread is created the control is returned back to the DMTCP and continues the execution of MPI application.

During the restart phase the control comes back to our plugin and the conditional\_variable is signaled. The thread waiting on the conditional\_variable now calls MPI\_Comm\_spawn and spawns additional required processes. MPI\_COMM\_WORLD is passed as the intracommunicator \cite{b15} to MPI\_Comm\_spawn while creating the additional processes. The child\_comm is used as intercommunicator to communicate with the newly spawned MPI processes. We now use MPI\_Intercomm\_merge to merge the child\_comm into resized\_world\_comm. The resized\_world\_comm is a intracommunicator which contains the old as well as the newly spawned MPI processes. We now assign resized\_world\_comm to MPI\_COMM\_RESIZED\_WORLD and relay the control back to the MPI application.

\begin{lstlisting}[caption=Code Snippet to create the resized world communicator, float]

/* intercommunicator to communicate 
 * with new spawned processes
 */ 
 
MPI_Comm child_comm; 
MPI_Comm_spawn("./worker", MPI_ARGV_NULL, 
                2, MPI_INFO_NULL, 0, 
                MPI_COMM_WORLD, 
                &child_comm, 
                MPI_ERRCODES_IGNORE);

MPI_Comm resized_world_comm;
MPI_Intercomm_merge(child_comm, 0, 
                    &resized_world_comm);

 MPI_COMM_RESIZED_WORLD = resized_world_comm;
\end{lstlisting}

This approach however has a caveat. The user MPI program needs to check in every iteration of the computation whether the MPI\_COMM\_RESIZED\_WORLD is not equal to MPI\_COMM\_NULL. As soon as this condition is satisfied the MPI old processes can perform the data re-partition to distribute the work amongst the newly spawned processes. This step is similar to receiving the MPI\_WORLD\_RESIZED error code in approach A. Listing 3 provides a code snippet to demonstrate the idea.

One can ask why not use the error code to indicate to the user about the resized world as described in approach A. The reason for this is to remain MPI implementation agnostic. This method does not require any changes in the MPI implementation. The implementation to increase the number of processes is completely encapsulated in the DMTCP plugin.

\subsection{Approach C}
Here, we formulate a new API to the MPI standard: MPI\_Fork. MPI\_Fork can have a wide applicability and prove to be a very useful tool to developers. One can argue there already exists MPI\_Comm\_spawn and MPI\_Comm\_spawn\_multiple which serves the purpose of spawning more processes. However the MPI processes started using MPI\_Comm\_spawn and MPI\_Comm\_spawn\_multiple have none of the parents data. The newly spawned processes needs to fetch the data to process from the other processes. This operation involves a use of intercommunicators \cite{b17} and these transforms a fairly simple MPI application into a complex one. 

MPI\_Fork tries to reduce this complexity by creating a clone of the parent and hence starting the child with already initialized parents data. The developers of MPI applications are usually scientists but not computer scientists. and introducing a simpler API of MPI\_Fork leads to its easy adoption amongst the community.

MPI\_Fork draws inspiration from UNIX fork for creating new processes. We propose using the same idea like UNIX to be able to fork new ranks from old ranks such that each new rank(child) forks off with the same memory of its parent. This way, we achieve new ranks that are initialized with some data i.e their parent's data. Hence we now the fully initialized data, and we can "forget" the part of the data that we don't need.

One fundamental difference between the UNIX fork and MPI\_Fork is that the UNIX forked process is spawned on the same machine. However the process created using MPI\_Fork may or may not be spawned on the same node as its parent rank.

The ability to fork a new child can allow sharing of responsibility. The newly forked processes can derive the mutually exclusive work based on who the parent is. Thus we bring the traditional parallelism model of forking can be made possible in MPI universe. One such example can be relaxation models e.g membrane of a soap bubble. The core idea here would be that at the beginning of the analysis we will have a coarser grid. Then on availability of new resources new children are forked. As the number of processes increases the grid becomes finer and finer and the accuracy of the model increases. Thus our multi-grid application would use limited resources at the beginning and would demand additional nodes only later.

The ability to increase the resources as the computation demands results into a more effective and optimal use of resources in computational clusters. Hellenbrand et. al. \cite{b16} demonstrates and concludes that the CPU hours to complete one simulation run is reduced almost half in the elastic MPI run.

\begin{lstlisting}[caption=Code Snippet for Approach B, float]

void repartition_data_if_required(){

	if (MPI_COMM_RESIZED_WORLD != MPI_COMM_NULL){
		/* code to repartition or 
		 * redistribute the data
		 */
		MPI_barrier();
	}
}

int main(){

	//...
	int i;
	for (i = 0; i < TOTAL_ITERATIONS; i++)
	{	
		/*
		 * user code
		 */
		repartition_data_if_required();
	}
	//...
}

\end{lstlisting}

We now propose the design for implementing MPI\_Fork in ExaMPI. We discussed the order of launching the Controller, Head and Fault daemons in the ExaMPI subsection of background. In order to implement MPI\_Fork we propose a design change in ExaMPI for the order in which the daemons are launched. We propose the following ordering where mpiexec launches the controller\_daemon. The controller\_daemon determines the number of head\_daemons required and launches the corresponding head\_daemons. The head\_daemons determined the number of fault\_daemons required and launches the corresponding fault\_daemons. 


The new design for MPI\_Fork necessitates changing the role of the fault\_daemons in launching other processes.
Recall that the fault\_daemons in the current design directly launches the MPI Application. In the proposed design, each fault\_daemon launches the MPI Application under DMTCP. This can be achieved by using dmtcp\_launch command provided by the DMTCP library. Each MPI process has its own DMTCP coordinator and the fault\_daemon tracks the information to communicate with its corresponding DMTCP coordinator.


There is a reason why we propose that each MPI process should be launched with separate instance of DMTCP. If we start the entire MPI cluster under a single instance of DMTCP then the entire cluster )along with Controller, Head and Fault daemons) will be checkpointed. The requirement of the implementation is to clone a single MPI process fundamentally which can only be achieved if each MPI process is started under a separate instance of DMTCP.


Next, we discuss the sequence of events that occur after a call to MPI\_Fork.
A call to MPI\_Fork signals the controller\_daemon. The controller\_daemon determines the MPI processes that should be cloned. This is determined by using the following deterministic algorithm. Since we have constraints on the the value of m, the controller\_daemon can select the first m ranks for the cloning. Once decided the fault\_daemons managing these MPI processes are signaled to checkpoint the corresponding MPI processes. This is achieved by using the dmtcp\_command --coord-host [host] --checkpoint. For the purpose of this paper we will consider the scenario of forking additional m processes. m can be between 1 and n, where n is the total number of MPI processes in the MPI application. Since MPI\_Fork is designed to be a collective operation, each MPI process should participate in it. 


After MPI\_Fork has signaled the controller\_daemon, the controller\_daemon must organize the remaining work.
The controller\_daemon determines the number of head\_daemons required based on the number of additional processes required. The controller\_daemon now launches the additional head\_daemons. The head\_daemons launches the corresponding number of fault\_daemons. The fault\_daemons launches or in this case restarts the corresponding checkpointed MPI processes. This is achieved using the dmtcp\_restart [checkpoint\_file]. During the restart process the environment variables and configs file of old processes are modified and connection information of the newly started processes is populated. This is to ensure the newly spawned processes are recognised by the old MPI processes. 

The proposed ordering of launching the daemons can also be applied to approach A. Approach A requires head\_daemon to  launch a fault\_daemon. Since the ordering between head\_daemon and fault\_daemon is same for both the approaches we can easily support both approach A and approach C using the proposed ordering in approach C.

While the new processes are being started the old MPI processes are waiting on a barrier which includes the parent as well the newly started processes. On successful restart the child processes now enter the barrier. Since the MPI processes are in the barrier, the control returns back to the MPI\_Fork in both old and the new MPI processes. In the old processes, MPI\_Fork returns a positive value which corresponds to number of new MPI processes spawned. In the new processes, MPI\_Fork returns zero. A negative return value indicates the error encountered during the forking process. 

For successful forking, an intracommunicator is required for peer-to-peer or collective communication across the old and new MPI processes. There are multiple options to solve this  problem. A very straight forward solution is to add all the newly forked to MPI\_COMM\_WORLD. Hence MPI\_COMM\_WORLD represents the entire universe of old and new MPI processes. There can arise a need where an operation needs to be performed on the processes which existed before forking. We will call those processes as the parents. Similarly a need can arise to address only the newly spawned processes.  We will call those processes as the children.


We introduce two new intracommunicators MPI\_COMM\_PARENTS and MPI\_COMM\_CHILDREN. MPI\_COMM\_PARENTS as the name suggests encapsulates the old processes or parent processes and MPI\_COMM\_CHILDREN encapsulates the newly spawned processes or the children processes. The initial value or the value before forking of MPI\_COMM\_PARENTS and MPI\_COMM\_CHILDREN would be MPI\_COMM\_NULL. However this solution does not consider the scenario where forking occurs repeatedly.

In order to solve the problem of multiple forking, we can use a versioned communicator as discussed in approach A. Each time MPI\_Fork is invoked, we add a version tag to all the three intracommunicators, viz. MPI\_COMM\_WORLD, MPI\_COMM\_PARENTS and MPI\_COMM\_CHILDREN. Now using the tag along with the communicator the MPI processes can easily distinguish between the processes which existed at various forking times.

\section{Discussion}
The approach A, in our opinion, is transparent to the user since it requires no modification to the user code, except for the part where redistribution of data after the world has been resized needs to be handled. However the disadvantage of approach A is it would require a rewrite of mpiexec and other \emph{fault\_daemons}. It also changes the way in which jobs are spawned in slurm. Now we would require to spawn a slurm job(\emph{fault\_daemon}) from within the slurm job(\emph{head\_daemon}). This might increase the complexity. Another disadvantage is that this approach is limited to ExaMPI and cannot be extended to other implementations of MPI.  

Approach B is cleaner and better than the Approach A since there is no need for any change in the underlying MPI implementation. However we completely rely on the APIs specified in MPI standard. However since ExaMPI MPI\_Comm\_spawn and MPI\_Intercomm\_merge are not supported in ExaMPI. And hence this approach cannot be used by application using ExaMPI.

Approach C introduces totally a new API and moreover a model to perform the computation. It brings MPI closer to the traditional fork and spawn model of programming and distributing the working amongst the child processes. 

With minimal change that do not alter the user code structure or logic, we showed that elastic MPI execution can be achieved. We showed that transformation of a tightly coupled parallel application into a malleable one is possible. This transformation could result into a better and resource efficient application. 

\section{Future Work}

Until now we discussed how to increase the MPI processes given the allocated resources increases. However there is also a scenario of reducing the allocated resources. In this case we need to reduce the number of MPI processes or migrate the existing MPI from the node which is supposed to freed up. Migration can be orchestrated using DMTCP and carefully selecting the process to migrate and restore it on the new node.


Another approach can be to have a event-sourcing model, where each MPI message is considered an event. Each event has sufficient information to identify the sender and receiver. Given that the MPI application goes through a process reduction phase, the event for the corresponding rank could be redirected to a proxy rank. A proxy rank is a live MPI process which gets elected by to send and receive events of the reduced process.

\section{Acknowledgements}
We thank Dr. Gene Cooperman \cite{b18} for their guidance throughout the course of this project and for their comments on the work and presentation. We wish to thank Dr. Anthony Skjellum \cite{b19} and the ExaMPI team for allowing us the access to ExaMPI source code and helping us understand the internals and architecture of ExaMPI.

\end{document}